\documentclass[aps,twocolumn,nofootinbib]{revtex4}

\usepackage{fancyhdr,fancybox}
\usepackage{hyperref}
\usepackage{wrapfig}
\usepackage{verbatim}
\usepackage{amsmath,amsfonts,amssymb}
\usepackage{url}
\usepackage[font=sf, labelfont={sf,bf}, margin=0.5cm]{caption}
\usepackage{graphicx}
\usepackage[euler]{textgreek}
\usepackage{adjustbox}
\usepackage{booktabs}
\usepackage{multirow}
\usepackage{xspace}
\usepackage{pdfpages}
\usepackage{stackengine}
\usepackage{textpos}
\usepackage{color, xcolor}

\newcommand\Bhinv{$\mathcal{B}_{H\rightarrow \mathrm{inv.}}$~}

\newcommand\hinv{H$\rightarrow$inv.~}
\newcommand\Hinv{H$\rightarrow$invisible~}
\newcommand\pt{$p_T$~}

\newcommand\ifb{fb$^{-1}$}

\fancypagestyle{firstPage}
{
	\pagestyle{fancy}
	\lhead{Amanda Steinhebel}
	\chead{}
}

\begin{document}

\title{H$\rightarrow$invisible at the ILC with SiD \\ \normalsize Talk presented at the International Workshop on Future Linear Colliders (LCWS2021), 15-18 March 2021. C21-03-15.1.}

\author{%
\textsc{Amanda Steinhebel, Jim Brau, Chris Potter}\\
\normalsize \textit {University of Oregon, Center for High Energy Physics, 1274 University of Oregon Eugene, Oregon 97403-1274 USA} \\
}
\date{\today}

\begin{abstract}
~\newline
\indent The Standard Model (SM) predicts a branching ratio of the Higgs boson decaying to invisible particles of $\mathcal{O}$(0.001), though current measurements have only set upper limits on this value. The small SM-allowed rate can be enhanced if the Higgs boson decays into new particles such as dark matter.  Upper limits have been placed on \Bhinv by ATLAS and CMS at $\mathcal{O}$(0.1), but the hadron environment limits precision. The ILC `Higgs factory' will provide unprecedented  precision of this electroweak measurement. Studies of the search for \Hinv processes in simulation are presented with SiD, a detector concept designed for the ILC. Preliminary results for expected sensitivity are provided, as well as studies considering potential systematics limitations.
\end{abstract}

\maketitle 


\section{Introduction}
\begin{figure}[ht!]
	\centering
	\includegraphics[width=0.7\linewidth]{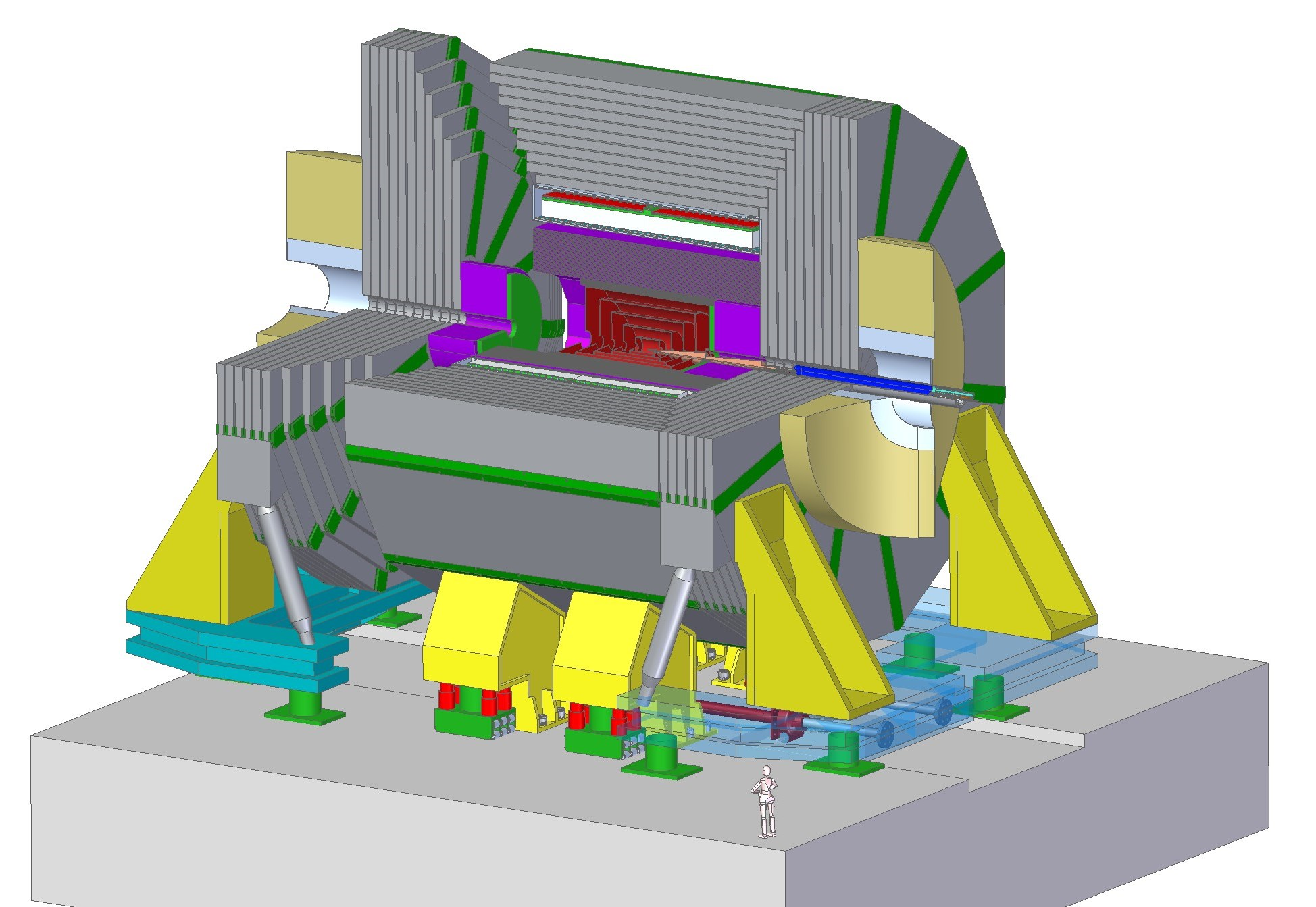}
	\caption{\label{fig:sid} The SiD detector \cite{tdr1} (Image: Marco Oriunno)}
\end{figure}

SiD is one of two detectors under consideration for use in the International Linear Collider (ILC) \cite{tdr4}. SiD, will feature high-granularity silicon tracking and calorimetry and is designed for particle flow reconstruction (see Figure~\ref{fig:sid}). SiD is a compact, cost-constrained, multi-purpose detector with silicon detectors used for vertexing, tracking, and electromagnetic calorimetry. These systems are housed inside a 5T field provided by a solenoid. 

There is a SM-predicted process ($H\rightarrow ZZ^*\rightarrow\nu\bar{\nu}\nu\bar{\nu}$) in which a Higgs boson ultimately decays to particles that are invisible to the detector - neutrinos. The branching ratio for this process is small, \Bhinv$\sim0.001$ \cite{hinvbr}, and difficult to measure. This process in particular carries potential implications for Higgs portal dark matter, where the addition of a Higgs decay to DM ($H\rightarrow\chi\chi$) would leave an identical detector signature. This addition could impact the expected branching ratio measurement by $\mathcal{O}(10\%)$ \cite{Kanemura_2010}. Any deviation of the measured branching ratio from the SM-only expectation could hint toward a DM contribution. 

The \hinv signature can be directly searched for by looking for events with missing energy from the invisible decay. The current strictest experimentally-set upper limit on \Bhinv{} using 139~\ifb{} of $\sqrt{s}=13$~TeV proton--proton collision data collected by the ATLAS detector at the LHC is 0.11 (0.11) observed (expected) \cite{combo}. This measured value is two orders of magnitude above the expected SM rate. Projected values estimate that the HL-LHC will be able to refine the observed upper limit to \Bhinv$<0.025$ \cite{hllhc}. The ILC Higgsstrahlung environment, providing selection of decay-independent Higgs decays, combined with the clean environment of the ILC Higgs Factory is expected to refine this value even further. The lack of hadronic interactions and pileup allows for more precise measurement of missing energy, measured with detectors designed for particle flow reconstruction. Updated studies using the ILD detector at the ILC250\footnote{ILC250 is the first ILC stage with electron-positron collisions at $\sqrt{s}=250$~GeV.} estimate the precision to measure \Bhinv$<0.0023$ \cite{ild}.

This document summarizes work done at the University of Oregon to investigate the ability of SiD to measure \Bhinv at the ILC250.

\section{Physics Overview and Strategy}

The ILC250 will create Higgs bosons predominantly through Higgsstrahlung, i.e. associated production of a $Z$ boson. The clean ILC environment allows for an indirect Higgs search technique to be employed. The visible decay of the $Z$ boson is first identified, through leptonic or hadronic decays. Once the $Z$ energy and mass are reconstructed, the Higgs can be identified as the invariant mass recoiling against the reconstructed $Z$ as
$$
	m_{recoil}^2=s+m_Z^2-2E_Z\sqrt{s}~,
$$
where $\sqrt{s}$ is the center of mass energy and $m_Z$ and $E_Z$ are the mass and energy of the reconstructed $Z$ boson. 

A search for \hinv{} utilizing Monte Carlo simulated data\footnote{These samples do not include beam effects such as beamstrahlung. Total initial yields in Tables~\ref{tab:sid_lep} and \ref{tab:sid_had} differ slightly due to considered Higgsstrahlung XS significant figures and minijet/1-fermion events considered in the full simulation used for the hadronic channel.} was conducted with a simulation of the SiD detector \cite{chris}. The studies presented here should be considered preliminary, and are under continued development.

For both $Z$ channels, a run scenario of 1800~\ifb{} is considered. The beams are polarized $(P_{e^-},P_{e^+}) = (\mp0.8, \pm0.3)$ with 900~\ifb{} collected for each polarization configuration\footnote{The polarization scheme for ILC250 plans to also include 100~\ifb{} of $(P_{e^-},P_{e^+}) = (\pm0.8, \pm0.3)$ as well, for a total of 2~ab$^{-1}$ \cite{ilc_polar}.}. Shorthand, these configurations are referred to by the particles left- or right-handedness (LR or RL).

Background sources are grouped by the number of fermions present in the final state, after boson decay. Backgrounds with 2-, 3-, and 4- fermions are relevant for these searches, as well as the inclusive SM decay of all Higgs produced through Higgsstrahlung. 

The branching ratio of the \hinv{} process is inflated to \Bhinv$=10\%$ for the plots and yields reported in this note. Yields of these signal events and the sum of background events for each kinematic requirement in the cutflow are used to calculate a significance,
$$
	\mathcal{S}=\frac{S}{\sqrt{S+B}}~,
$$
where $S$ and $B$ are signal and background yields, respectively. From this significance, a rough estimate of the upper limit can be calculated where 
$$
	95\%~\mathrm{UL} [\%] =\frac{\mathrm{SM}~\mathcal{B}_{H\rightarrow \mathrm{inv.}}[\%]\times1.65}{\mathcal{S}}~.
$$

\section{Leptonic $Z$ Channel}

Dominant backgrounds for the leptonic $Z$ channel include 2-fermion processes ($e^+e^-\rightarrow\mu^+\mu^-$) and 4-fermion processes ($e^+e^-\rightarrow ZZ\rightarrow\nu\bar{\nu}\ell\ell$ and $e^+e^-\rightarrow WW\rightarrow\ell\nu\ell\nu$). All 2-, 3-, and 4- fermion background processes are considered, as well as inclusive SM Higgs decays. Contributing SM $H$ processes are dominated by $H\rightarrow bb$ and $H\rightarrow WW^*$ with additional contributions from $H\rightarrow ZZ^*$, $H\rightarrow cc$, and $H\rightarrow\tau\tau$. Samples are generated with the SiD \cite{dsid} fast simulation in Delphes \cite{delphes1, delphes2} with lepton isolation\footnote{Lepton isolation requirements are made for particle ID at the simulation stage, rather the analysis stage. Here, all leptons pass lepton isolation criteria - 12\% (15\%) relative $E_\mathrm{T}$~(\pt) contained in a cone of radius $R=0.5$ around the reconstructed electron (muon).}.

Lepton requirements are made to isolate and identify the visible $Z$ decays. This requires two same-flavor-opposite-sign (SFOS) leptons with an invariant mass within 20~GeV{} of the $Z$ mass and visible \pt{} between $20-70$~GeV. The recoil mass must fall within a Higgs mass window, $110<m_{\mathrm{recoil}}<150$~GeV. A MET\footnote{Missing transverse energy (MET) is used.} requirement of MET$>15$~GeV is also required (see Table~\ref{tab:sid_lep}). Beam polarization greatly impacts the prevalence of different background sources (see Figure~\ref{fig:sid_lep_mass}).

\begin{table*}[ht]
	\centering
	\begin{tabular}{|l||c|c|c||c|c|c|}\hline
		Cut& S (LR)&B (LR)&$\mathcal{S}$&S (RL)&B (RL) & $\mathcal{S}$\\\hline
		All events                     & 2.82e4  & 1.32e8 & 2.45 & 1.90e4 & 5.85e7 & 2.48 \\
		MET$>15$ GeV                   & 2.02e4  & 3.42e7 & 3.46 & 1.43e4 & 5.80e6 & 5.94 \\
		2 leptons                      & 1.46e3  & 1.61e6 & 1.15 & 1.04e3 & 1.97e5 & 2.33 \\
		SFOS leptons $>10$ GeV         & 1.38e3  & 1.37e6 & 1.18 & 984    & 1.39e5 & 2.63 \\
		$75<$M\_vis$<105$ GeV          & 1.31e3  & 2.92e5 & 2.42 & 933    & 5.81e4 & 3.84 \\
		$20<$pt\_vis$<70$ GeV          & 1.27e3  & 2.15e5 & 2.73 & 905    & 3.98e4 & 4.48 \\
		$110<$m\_recoil$<150$ GeV      & 1.25e3  & 1.13e5 & 3.71 & 892    & 1.62e4 & 6.82 \\\hline

	\end{tabular}
	\caption{\label{tab:sid_lep} Cutflow for \hinv{} search with leptonic $Z$ decays}
\end{table*}

\begin{figure}[ht!]
	\centering
	\includegraphics[width=\linewidth]{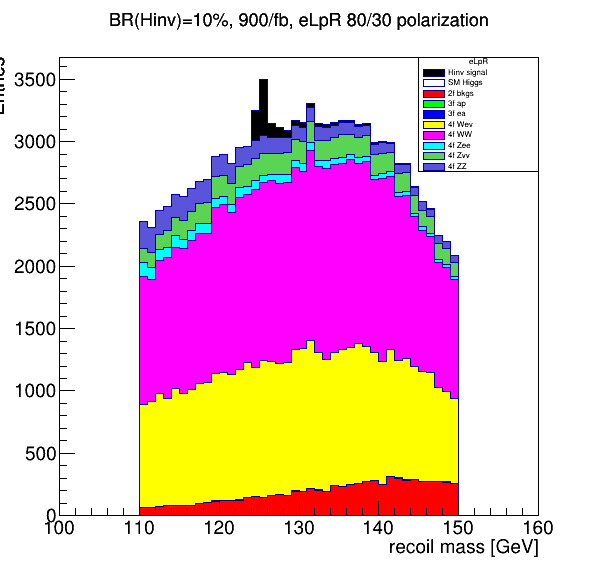}\\
	\includegraphics[width=\linewidth]{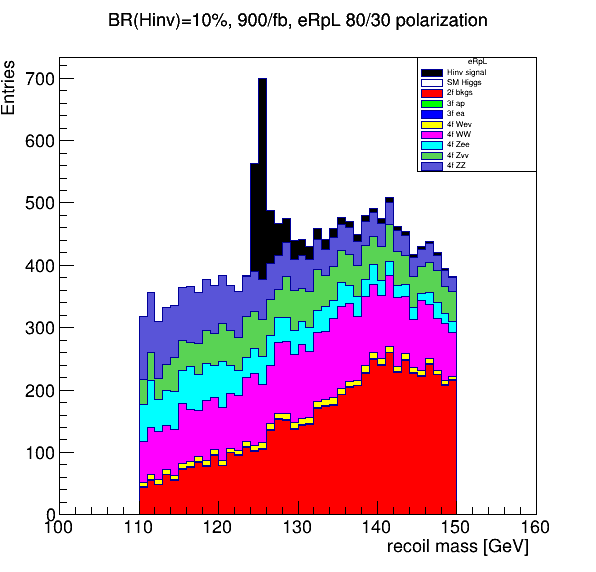}
	\caption{\label{fig:sid_lep_mass} Recoil mass distributions of events passing all kinematic criteria of Table~\ref{tab:sid_lep}}
\end{figure}

\section{Hadronic $Z$ Channel}

Dominant backgrounds for the leptonic $Z$ channel include 4-fermion ($e^+e^-\rightarrow ZZ\rightarrow\nu\bar{\nu}q\bar{q}$ and $e^+e^-\rightarrow WW\rightarrow\ell\nu q\bar{q}$) and 2-fermion processes ($e^+e^-\rightarrow q\bar{q}$). 3-fermion backgrounds play a larger role in this channel, accounting for roughly 30\% of all background events. All 2-, 3-, and 4- fermion background processes are considered\footnote{Inclusive SM Higgs decays are not included.}. Samples are generated with the SiD full simulation in ILCSoft \cite{ilcsoft} v02-00-02 with SiD option 2 version 3.

Jet requirements are made to isolate and identify the visible $Z$ decays. This requires that the event contain no leptons and exactly two jets with an invariant mass within 20~GeV{} of the $Z$ mass and visible \pt{} between $20-70$~GeV. The recoil mass must fall within a Higgs mass window, $110<m_{\mathrm{recoil}}<150$~GeV. When possible, identical selections as the leptonic channel are chosen (see Table~\ref{tab:sid_had}). The lepton veto leaves the two channels orthogonal.

\begin{table*}[ht]
	\centering
	\begin{tabular}{|l||c|c|c||c|c|c|}\hline
		Cut& S (LR)&B (LR)&$\mathcal{S}$&S (RL)&B (RL) & $\mathcal{S}$\\\hline
		All Events & 2.79e4 & 2.6e9 & 0.547 & 1.89e4 & 2.54e9 & 0.375 \\
		$N_{e}+N_{\mu}=0$ & 2.17e4 & 2.15e9 & 0.469 & 1.45e4 & 2.11e9 & 0.316 \\
		$N_{trk} \in [6,24], N_{pfo} \in [12,40]$ & 1.36e4 & 1.5e8 & 1.11 & 9.67e3 & 1.2e8 & 0.884 \\
		$20 \leq p_{T}^{vis} \leq 60$~GeV & 1.25e4 & 7.71e6 & 4.48 & 8.84e3 & 1.07e6 & 8.53 \\
		$75 \leq m_{vis} \leq 105$~GeV & 1.16e4 & 1.79e6 & 8.63 & 8.21e3 & 3.14e5 & 14.5 \\
		$N_{jet}=2$ & 1.16e4 & 1.79e6 & 8.63 & 8.21e3 & 3.14e5 & 14.5 \\
		$-0.9 \leq \cos \theta_{jj} \leq -0.2$ & 1.08e4 & 8.68e5 & 11.5 & 7.65e3 & 1.78e5 & 17.7 \\
		$110 \leq m_{recoil} \leq 140$~GeV & 1.03e4 & 3.6e5 & 17 & 7.33e3 & 8.39e4 & 24.2 \\\hline
	\end{tabular}
	\caption{\label{tab:sid_had} Cutflow for \hinv{} search with hadronic $Z$ decays}
\end{table*}

Further optimization of this channel is underway utilizing multivariate analysis techniques to improve signal selection. Preliminary studies show up to a factor of two improvement of the upper limit of \Bhinv.

\section{Combination}

Since the leptonic and hadronic channels are orthogonal, they can be simply considered in combination (see Table~\ref{tab:sid_combo}). The highest significance, found by combining $Z$ decay channels and polarization schemes, leads to an estimated \Bhinv$\lesssim0.0054$. This is comparable to \Bhinv$<0.0036$ at 95\% C.L.\footnote{Assuming SM Higgs total width}, estimated with a 2~ab$^{-1}$ data set modeled with the expanded polarization scheme \cite{ilc_polar}. The SiD result can be improved by understanding systematic uncertainties (see next section), using full simulation rather than the fast simulation utilized in this study, expanding to include the full 2~ab$^{-1}$ polarization scheme, and optimizing kinematic cut requirements. Nevertheless, this is two orders of magnitude stricter than the current LHC best limit \Bhinv$<0.11(0.11)$ and one order of magnitude stricter than the anticipated HL-LHC performance of \Bhinv$\lesssim0.025$. ILD studies of ILC250 similarly estimate the ability to measure \Bhinv$<0.0023$ \cite{ild}.

\begin{table}[ht!]
	\centering
	\begin{tabular}{|l|l|c|c|c|c|}\hline
		~&~&S Yield &B Yield&$\mathcal{S}$&BR U.L. [\%]\\\hline
		\multirow{2}{*}{$Z$(had)}&eLpR&1.03e4&3.6e5&16.9&0.97\\
		&eRpL&7.33e3&8.39e4&24.2&0.68\\\hline
		\multirow{2}{*}{$Z$(lep)}&eLpR&1.25e3&1.13e5&3.71&4.45\\
		&eRpL&892&1.62e4&6.82&2.42\\\hline
		\multirow{2}{*}{Combined}&eLpR&1.16e4&4.73e5&17.3&0.95\\
		&eRpL&8.22e3&1.00e5&25.1&0.65\\
		&Combined&1.98e4&5.73e5&30.5&0.54\\\hline
	\end{tabular}
	\caption{\label{tab:sid_combo} Combined final yields of Tables~\ref{tab:sid_lep} and \ref{tab:sid_had} for 1800~\ifb{} of LR/RL polarized beams. Significance values for combined channels are calculated from quadrature sum of inputs. }
\end{table}

\section{Systematic Uncertainties}

A number of studies of uncertainties are actively progressing. One goal is to understand which systematic errors may drive the degradation of the expected upper limit on \Bhinv, and mitigate the effects if possible. It is known, for example, that jet energy resolution plays a large role. This must be quantified. Minor impacts on resolution from jet flavor (decay products of the $Z$) have also been noted (see Figure~\ref{flavor}). 

\begin{figure}[ht!]
	\centering
	\includegraphics[width=\linewidth]{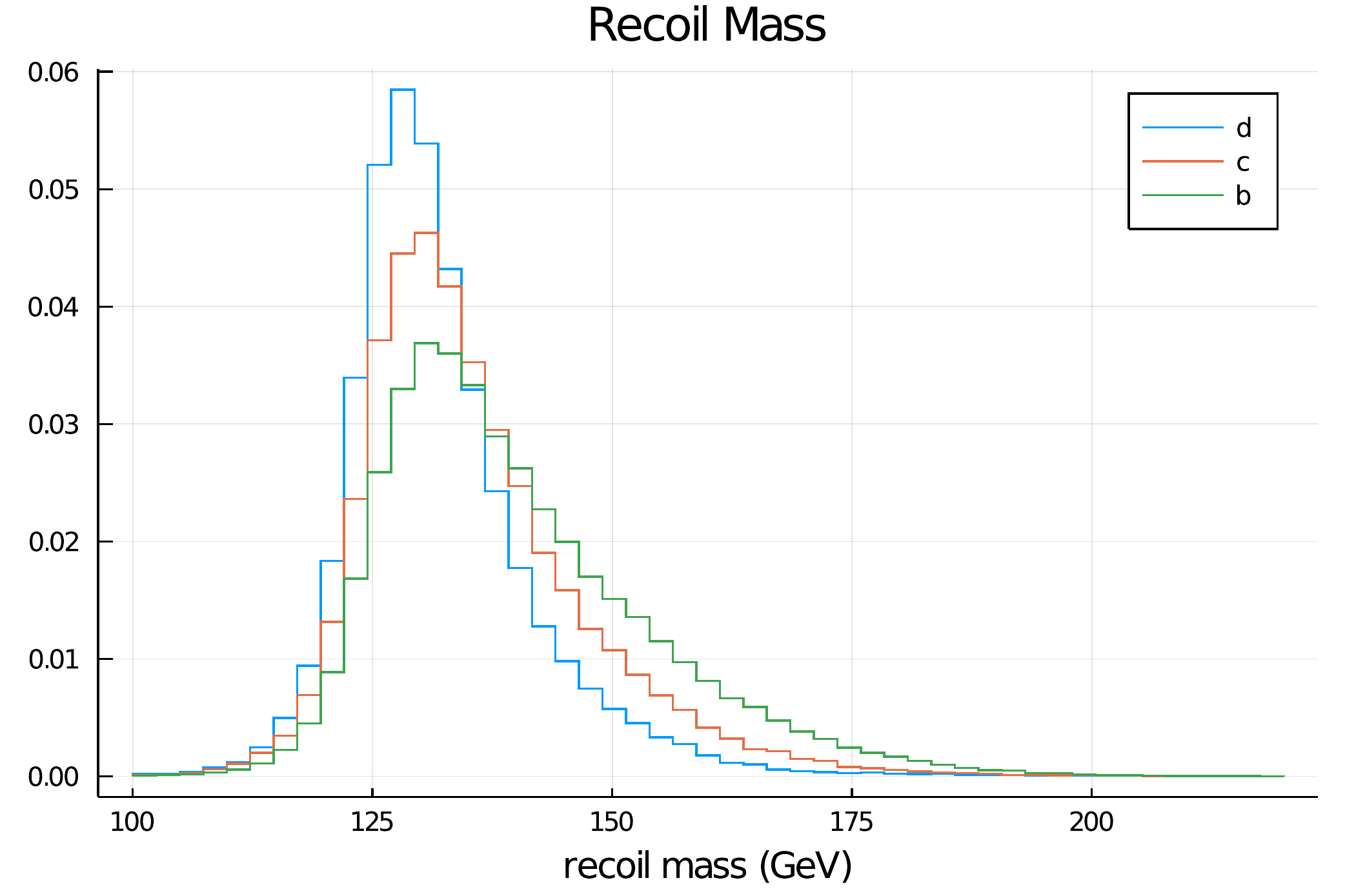}
	\caption{\label{flavor} Recoil mass distributions from heavy-flavor $Z$ decays}
\end{figure}

Studies are also underway to understand the impact that the SiD geometry may play on measuring the \hinv{} process, and to understand the strengths of the SiD design. The Consortium is considering the impact of hadronic calorimeter tile size on energy resolution, and investigating the inclusion of Monolithic Active Pixels (MAPs) for silicon readout. 

\section{Summary}

The SiD detector at the ILC is expected to measure \Bhinv{} more than 100 times more precisely that that which has been done to date at the LHC. This document overviews a cut-and-count method using 1800~\ifb{} of $\sqrt{s}=250$GeV $e^+e^-$ data with $(P_{e^-},P_{e^+}) = (\mp0.8, \pm0.3)$ to estimate how the SiD detector may measure \Bhinv. By combining polarizations and leptonic and hadronic $Z$ decay channels from Higgsstrahlung events, it is seen that SiD could measure \Bhinv$\lesssim0.0054$. This is comparable to ILD studies considering the full 2~ab$^{-1}$ of polarized data anticipated to be collected. 

Work remains underway to further refine this performance with detector improvements and multivariate analysis techniques, and to understand limitations of systematic uncertainties.

\section{Acknowledgments}

We would like to thank Jan Strube, Andy White, Austin Prior, Marty Breidenbach, and Makayla Massar for their contributions to this work.


\end{document}